\documentclass{article}
\usepackage[export]{adjustbox}
\usepackage{spconf,amsmath,graphicx}

\usepackage[utf8]{inputenc} 
\usepackage[T1]{fontenc}    
\usepackage{hyperref}       
\usepackage{url}            
\usepackage{booktabs}       
\usepackage{amsfonts}       
\usepackage{nicefrac}       
\usepackage{microtype}      
\usepackage{amssymb}

\def\lb{\label}
\def\erf#1{(\ref{#1})}

\def\bSigma{\boldsymbol{\Sigma}}

\def\bb{\mbox{$\mathbf{b}$}}

\def\bX{\mbox{$\mathbf{X}$}}

\def\bv{\mbox{$\mathbf{v}$}}

\def\bw{\mbox{$\mathbf{w}$}}
\def\bwH{\mbox{$\mathbf{w}$}^H}
\def\bvH{\mbox{$\mathbf{v}$}^H}

\def\bp{\mbox{$\mathbf{p}$}}

\def\bzero{\mbox{$\mathbf{0}$}}

\def\SigN{\mbox{$\Sigma_{\mathbf{N}}$}}

\long\def\gobbleup#1{}

\def\bSigmaNi{\mbox{$\bSigma_{\mathbf{N}}^{-1}$}}

\def\fidx{(\omega_k)}
\def\fpidx{(\omega_k,\bp)}
\def\tfidx{(t,\omega_k)}
\def\tfpidx{(t,\omega_k,\bp)}

\def\RE{\mbox{$\ \operatorname{Re}$}}
\def\IM{\mbox{$\ \operatorname{Im}$}}

\title{Frequency Domain Multi-channel Acoustic Modeling for Distant Speech Recognition}
\name{Minhua Wu, Kenichi Kumatani, Shiva Sundaram, 
Nikko Str{\"{o}}m, Bj{\"{o}}rn Hoffmeister\thanks{We would like to acknowledge the support of our colleagues, Arindam Mandal, Brian King, Gautam Tiwari, I-Fan Chen, Jeremie Lecomte, Lucas Seibert, Roland Maas, Sergey Didenko and Zaid Ahmed.}}
\address{Amazon Inc.}
\begin{document}
\ninept
\maketitle
\begin{abstract}
Conventional far-field automatic speech recognition (ASR) systems typically employ microphone array techniques for speech enhancement in order to improve robustness against noise or reverberation. However, such speech enhancement techniques do not always yield ASR accuracy improvement because the optimization criterion for speech enhancement is not directly relevant to the ASR objective. In this work, we develop new acoustic modeling techniques that optimize spatial filtering and long short-term memory (LSTM) layers from multi-channel (MC) input based on an ASR criterion directly. In contrast to conventional methods, we incorporate array processing knowledge into the acoustic model. Moreover, we initialize the network with beamformers' coefficients. We investigate effects of such MC neural networks through ASR experiments on the real-world far-field data where users are interacting with an ASR system in uncontrolled acoustic environments. We show that our MC acoustic model can reduce a word error rate (WER) by~16.5\% compared to a single channel ASR system with the traditional log-mel filter bank energy (LFBE) feature on average. Our result also shows that our network with the spatial filtering layer on two-channel input achieves a relative WER reduction of~9.5\% compared to conventional beamforming with seven microphones. 
\end{abstract}
\section{Introduction}
\lb{sec:intro}
A complete system for distant speech recognition (DSR) typically consists of distinct components such as a voice activity detector, speaker localizer, dereverberator, beamformer and acoustic model~\cite{Omolog2001,Wolfel2009,KumataniAYMRST12,KinoshitaDGHHKL16,VirtanenBook2012}. Beamforming is a key component in DSR. Such beamforming techniques can be categorized into fixed and adaptive beamforming. The fixed beamforming (FBF) design provides better recognition accuracy than single microphone systems in many DSR applications. However, its noise suppression performance is often limited because of a mismatch between theoretical and actual noise field assumptions. In order to overcome such an issue, adaptive beamforming (ABF) or blind source separation techniques had been also applied to DSR tasks. The ABF techniques have been proven to improve noise robustness by using a dereverberation technique~\cite{Delcroix2014} or together with higher-order statistics~\cite{KumataniAYMRST12}. Those ABF methods normally rely on accurate voice activity detection~\cite{KinoshitaDGHHKL16}, mask estimation~\cite{Heymann2018} or speaker location performance~\cite[\S10]{Wolfel2009}. It is generally very challenging to identify voice activity from a desired speaker or track the target speaker in many DSR scenarios~\cite{Fiscus2008}. If such information is not obtained reliably, conventional ABF methods largely degrade recognition performance than FBF~\cite{Sullivan96,HimawanMS11}. While it is tempting to isolate and optimize each component individually, experience has proven that such an approach cannot lead to optimal performance~\cite{McDonough2008,Seltzer2008}. 

A straightforward approach to solving this problem would be simultaneously optimizing an audio processing front-end and acoustic model based on an ASR criterion. This approach was first pursed with Gaussian mixture model-based hidden Markov model (GMM-HMM)~\cite{Seltzer2004,Rauch2008}. However, due to the limited scalability of a linear beamforming model, the model has to be adapted for each acoustic environment every time, which makes real-time implementation hard. 

Such an adaptation process may not be necessary when we train a deep neural network (DNN) with a large amount of multi-channel (MC) data. It is also straightforward to jointly optimize the unified MC DNN so as to achieve better discriminative performance of acoustic units~\cite{Xiao16,Sainath2017,OchiaiICML17}. 
Those unified MC DNN techniques can fall into the following categories: 1) mapping a time-delay feature to oracle beamformer's weight computed with the ground-truth source direction through the DNN~\cite{Xiao16}, 2) feeding MC speech features into the network such as the log energy-based features~\cite{SwietojanskiSPL14,Braun2018} or LFBE supplemented with the time delay feature~\cite{KimInterspeech16,FujimotInterspeech17}, 3) applying convolutional neural networks to the MC signal in the time domain~\cite{Sainath2017} and 4) transforming the MC frequency input with the complex linear projection~\cite{Sainath2017,LiSNCBMSSPCSWWV17}. 
The performance of those methods would be limited due to the lack of the proper sound wave propagation model. As it will be clear in section~\ref{sec:conventional_system}, the DNN can subsume multiple fixed beamformers. Notice that the feature extraction components described in~\cite{Xiao16,SwietojanskiSPL14,Braun2018,KimInterspeech16} are not fully learnable. 

The unified MC acoustic modeling approach can provide a more optimum solution for the ASR task but requires a larger amount of training data for better generalization. Alternatively, a neural network can be also applied to the clean speech reconstruction task explicitly~\cite{Heymann2018}. Heymann et al. and Erdogan et al. proposed an LSTM mask method that estimated statistics of target and interference signals for ABF~\cite{Heymann2018,ErdoganInterspeech16} and MC Wiener filtering~\cite{Heymann2018}. It was further extended to an end-to-end framework by jointly optimizing the beamformer and attention-based encoder-decoder with the character error rate (CER) criterion~\cite{OchiaiICML17}. However, it should be noted that the mask-based beamforming technique needs to accumulate statistics from a certain amount of adaptation data or whole utterance data in order to maintain the improvement~\cite{Heymann2018,Higuchi2018}. Due to necessity of accumulating the sufficient statistics, it may cause a noticeable latency undesirable for real-time applications such a speech dialogue system.

In this work, we focus on development of fully learnable MC acoustic modeling. We consider three types of MC network architectures: complex affine transform, deterministic spatial filter selection with max-pooling, and elastic spatial filtering combination. The latter two architectures incorporate array processing knowledge into the MC input layer. All the neural networks use frequency input for the sake of computational efficiency~\cite{Haykin2001}. We evaluate those techniques on the real-world far-field data spoken by thousands of real users, collected in various acoustic environments. Therefore, the test data contains challenging conditions where speakers interact with the ASR system without any restriction under reverberant and noisy environments. 


\vspace{-0.25em}
\section{Conventional DSR System}
\lb{sec:conventional_system}
\vspace{-0.25em}

\subsection{Acoustic Beamforming}
\lb{sec:beamforming}
Let us assume that a microphone array with $M$ sensors captures a sound wave propagating from a position and denote the frequency-domain snapshot as $\bX \tfidx =[X_1 \tfidx,\cdots,X_{M} \tfidx]^T$ for an angular frequency $\omega_k$ at frame $t$. With the complex weight vector for source position $\bp$
\vspace{-0.5em}
\begin{equation}
\bw \tfpidx = [ w_1 \tfpidx, \cdots, w_{M} \tfpidx ] ,
\lb{eq:bfw}
\vspace{-0.5em}
\end{equation}
the beamforming operation is formulated as
\vspace{-0.5em}
\begin{equation}
Y \tfpidx = \bwH \tfpidx \bX \tfidx, 
\lb{eq:bfo}
\vspace{-0.5em}
\end{equation}
where $H$ is the Hermitian (conjugate transpose) operator. The complex vector multiplication~\erf{eq:bfo} can be also expressed as the real-valued matrix multiplication:
\vspace{-0.5em}
\begin{align}
\begin{bmatrix}
           \RE (Y) \\
           \IM (Y) \\
\end{bmatrix}
           &=
\begin{bmatrix}
         \RE (w_{1}) & \IM (w_{1}) \\ 
        -\IM (w_{1}) & \RE (w_{1}) \\
         \vdots &  \vdots \\
         \RE (w_{M}) & \IM (w_{M}) \\
        -\IM (w_{M}) & \RE (w_{M})\\
\end{bmatrix}^T
\begin{bmatrix}
           \RE (X_{1}) \\
           \IM (X_{1}) \\
           \vdots \\
           \RE (X_{M}) \\
           \IM (X_{M}) \\
\end{bmatrix},
\lb{eq:cat2}
\end{align}
where $\tfpidx$ is omitted for the sake of simplicity. It is clear from~\erf{eq:cat2} that beamforming can be implemented by generating $K$ sets of $2 \times 2M$ matrices where $K$ is the number of frequency bins. Thus, we can readily incorporate this beamforming framework into the DNN in either the complex or real-valued form. Notice that since our ASR task is classification of acoustic units, the real and imaginary parts in~\erf{eq:cat2} can be treated as two real-valued feature inputs. In a similar manner, the real and imaginary parts of hidden layer output can be treated as two separate entities. In that case, the DNN weights can be computed with the real-valued form of the back propagation algorithm\footnote{Although Haykin noted in~\cite[S17.3]{Haykin2001} that the convergence performance could degrade due to unnecessary degree of freedom to solve the complex mapping problem when a complex valued weight was treated as independent parameters, we have not observed any noticeable difference in our experiments. Thus, we treat the complex weight as independent entities unless we explicitly state that network has the complex affine transform.}.

A popular method in the field of ASR would be super-directive (SD) beamforming that uses the \emph{spherically isotropic noise field}~\cite{DocloM07,HimawanMS11}~\cite[S13.3.8]{Wolfel2009}. Let us first define the $(m,n)$-th component of the spherically isotropic noise coherence matrix as
\vspace{-0.25em}
\begin{equation}
\SigN_{m,n} \fidx = \text{sinc} \left( \omega_k d_{m,n} / c \right)
\lb{eq:NCM}
\vspace{-0.5em}
\end {equation}
where $d_{m,n}$ is the distance between the $m$-th and $n$-th sensors and $c$ is speed of sound.  This represents the spatial correlation coefficient between the $m$-th and $n$-th sensor inputs in the spherically isotropic noise (diffuse) field. The weight vector of the SD beamformer can be expressed as
\vspace{-0.5em}
\begin{equation}
\bwH_{\text{SD}} = \left[ \bvH \bSigmaNi \bv \right]^{-1} \bvH \bSigmaNi
\lb{eq:SD1}
\vspace{-0.5em}
\end {equation}
where \( \fpidx \) are omitted and $\bv$ represents the array manifold vector for time delay compensation. In order to control white noise gain, diagonal loading is normally adjusted~\cite[S13.3.8]{Wolfel2009}.

Although speaker tracking has a potential to provide better performance~\cite[\S10]{Wolfel2009}, the simplest solution would be selecting a beamformer based on normalized energy from multiple instances with various look directions~\cite{HimawanMS11}. In our preliminary experiments, we found that competitive speech recognition accuracy was achievable by selecting a fixed beamformer with the highest energy. Notice that highest-energy-based beamformer selection can be readily implemented with a max-pooling layer as described in section \ref{sec:MCDNN}. 

\subsection{Acoustic Model with Signal Processing Front-End}
\lb{sec:baseline}

As shown in figure~\ref{fig:dnn_baseline}, the baseline DSR system consists of an audio signal processing, speech feature extraction and classification NN components. The audio signal processing front-end transforms a time-discrete signal into the frequency domain and select the output from one of multiple beamformers based on the energy criterion. After that, the time-domain signal is reconstructed and fed into the feature extractor. The feature extraction step involves LFBE feature computation as well as causal and global mean-variance normalization~\cite{King2017}. The NN used here consists of multiple LSTM, affine transform and softmax layers. The network is trained with the normalized LFBE features in order to classify senones associated with the HMM state. In the conventional DSR system, the signal processing front-end can be separately tuned based on empirical knowledge. However, it may not be straightforward to jointly optimize the signal processing front-end and classification network~\cite{Heymann2018}, which will result in a suboptimal solution for the senone classification task.
\begin{figure}[t]
\addtolength{\belowcaptionskip}{-3.0em}
\begin{minipage}[t]{0.45\linewidth}
 \includegraphics[width=0.9\linewidth]{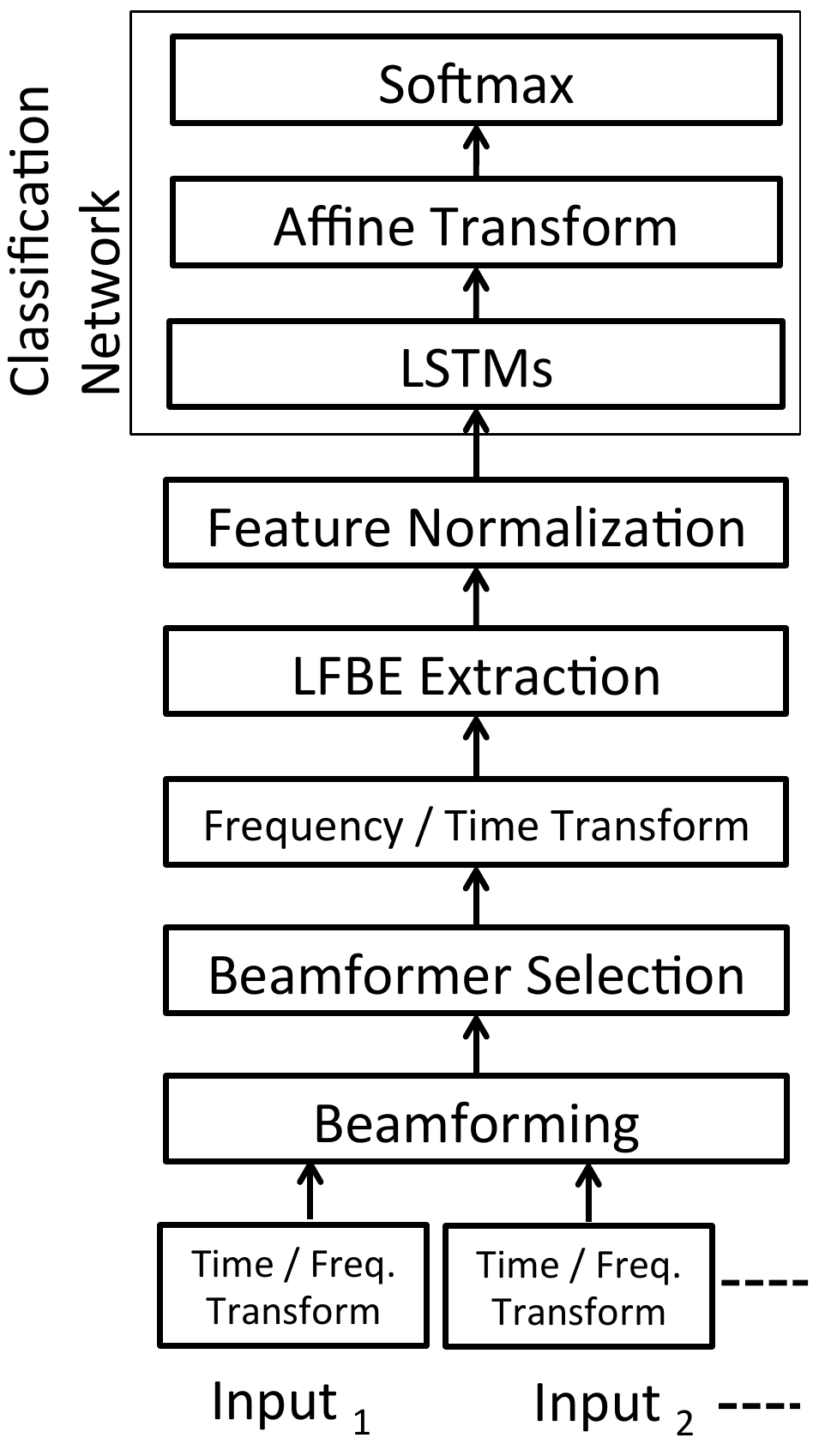}
 \vspace{-0.75em}
 \caption{Conventional system}
 \lb{fig:dnn_baseline}
\end{minipage}
\begin{minipage}[t]{0.05\linewidth}
\includegraphics[width=\linewidth]{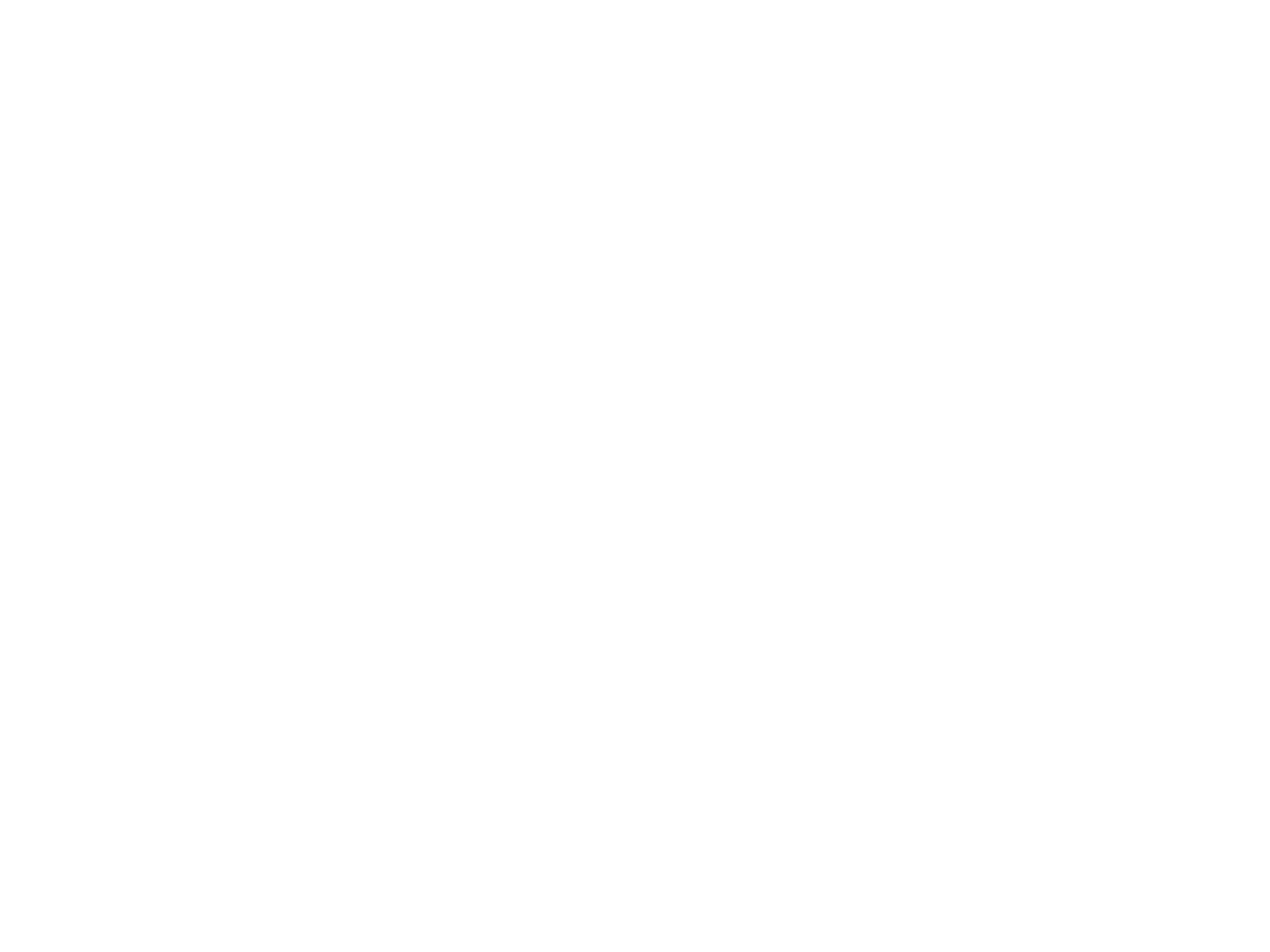}
\end{minipage}
\begin{minipage}[t]{0.45\linewidth}
 \includegraphics[width=0.9\linewidth]{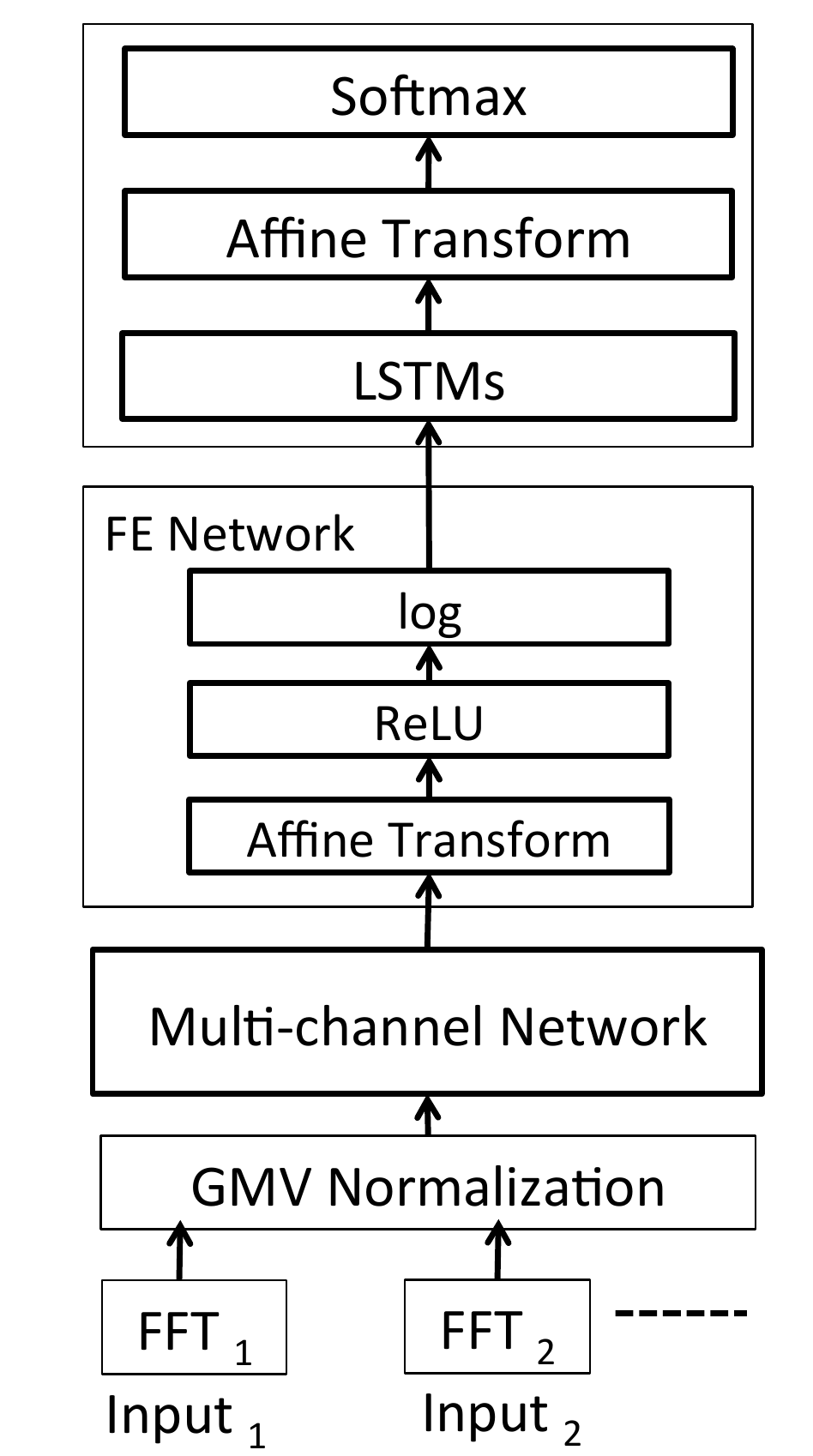}
 \vspace{-0.75em}
 \caption{Fully-learnable system}
 \lb{fig:dnn_main}
\end{minipage}
\vspace{-1.5em}
\end{figure}

\vspace{-0.25em}
\section{Frequency Domain Multi-channel Network}
\lb{sec:MCDNN}
\vspace{-0.25em}

As it was clear in section~\ref{sec:conventional_system}, conventional beamforming solutions are suboptimal for the speech recognition task. In this section, we describe our multi-channel (MC) network architecture that can be jointly optimized.  

Figure~\ref{fig:dnn_main} shows our whole DSR system with the fully-learnable neural network. As shown in figure~\ref{fig:dnn_main}, our DSR consists of 4 functional blocks, signal pre-processing, MC DNN, feature extraction (FE) DNN and classification LSTM. First, a block of each channel signal is transformed into the frequency domain through FFT. In the frequency domain, DFT coefficients are normalized with global mean and variance estimates. The normalized DFT features are concatenated and passed to the MC DNN. Our FE DNN contains an affine transform initialized with mel-filter bank values, rectified linear unit (ReLU) and log component. Here, the ReLU component is used in order to avoid putting a negative number into the log function. Notice that the initial FE DNN mimics the LFBE feature. The output of the FE DNN is then input to the same classification network architecture as the LFBE system, LSTM layers followed by affine transform and softmax layers. The DNN weights are trained in a stage-wise manner~\cite{Kumatani2017}; we first built the classification LSTM with the single channel LFBE feature, then trained the cascade network of the FE and classification layers with the single-channel DFT feature, and finally performed joint optimization on the whole network with MC DFT input. In contrast to the conventional DSR system, this fully learnable acoustic model approach neither requires clean speech signal reconstruction nor perceptually-motivated filter banks~\cite{RichardSN13}. 

\begin{figure*}[t]
\addtolength{\belowcaptionskip}{-2em}
\begin{minipage}[c]{0.66\linewidth}
\begin{minipage}[t]{0.31\linewidth}
\includegraphics[width=\linewidth]{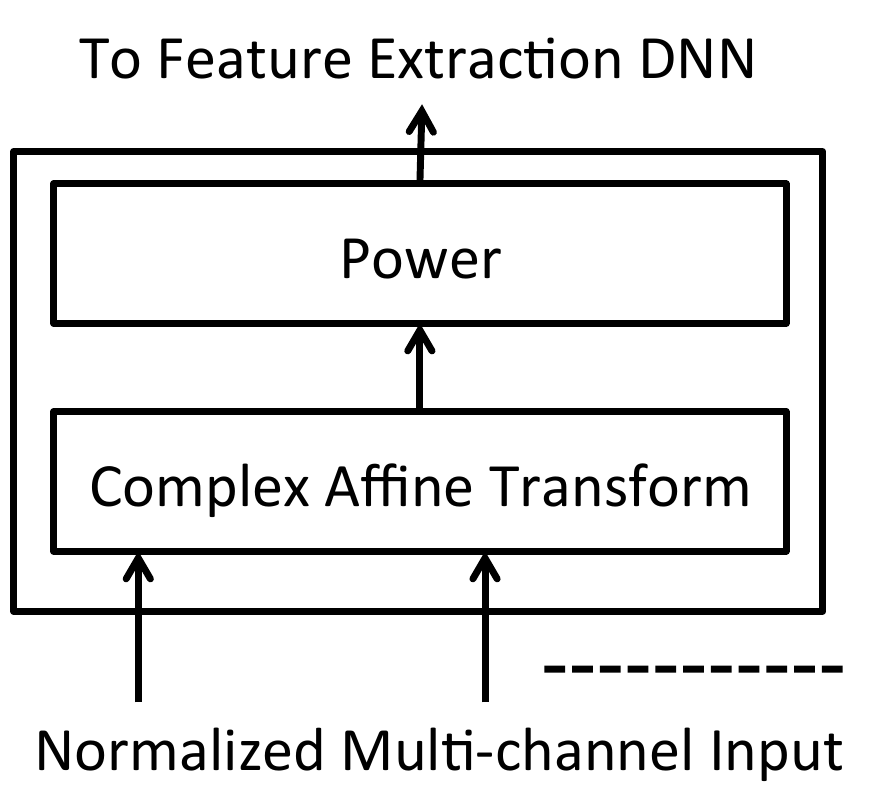}
\vspace{-0.75em}
\centering
\text{(a) Complex affine transform}
\lb{fig:mcdnn1}
\end{minipage}
\begin{minipage}[t]{0.01\linewidth}
\includegraphics[width=\linewidth]{white.pdf}
\end{minipage}
\begin{minipage}[t]{0.31\linewidth}
\includegraphics[width=\linewidth]{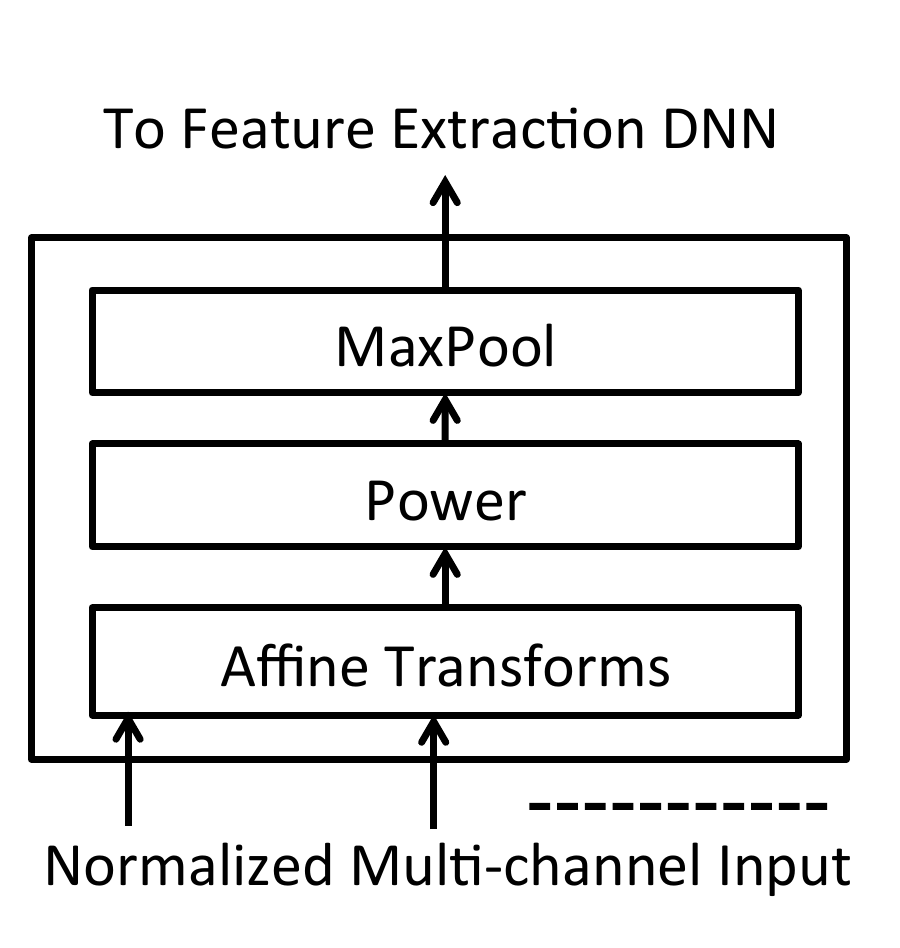}
\vspace{-0.75em}
\centering
\text{(b) Deterministic SF }
\lb{fig:det-mcdnn}
\end{minipage}
\begin{minipage}[t]{0.01\linewidth}
\includegraphics[width=\linewidth]{white.pdf}
\end{minipage}
\begin{minipage}[t]{0.31\linewidth}
\includegraphics[width=\linewidth]{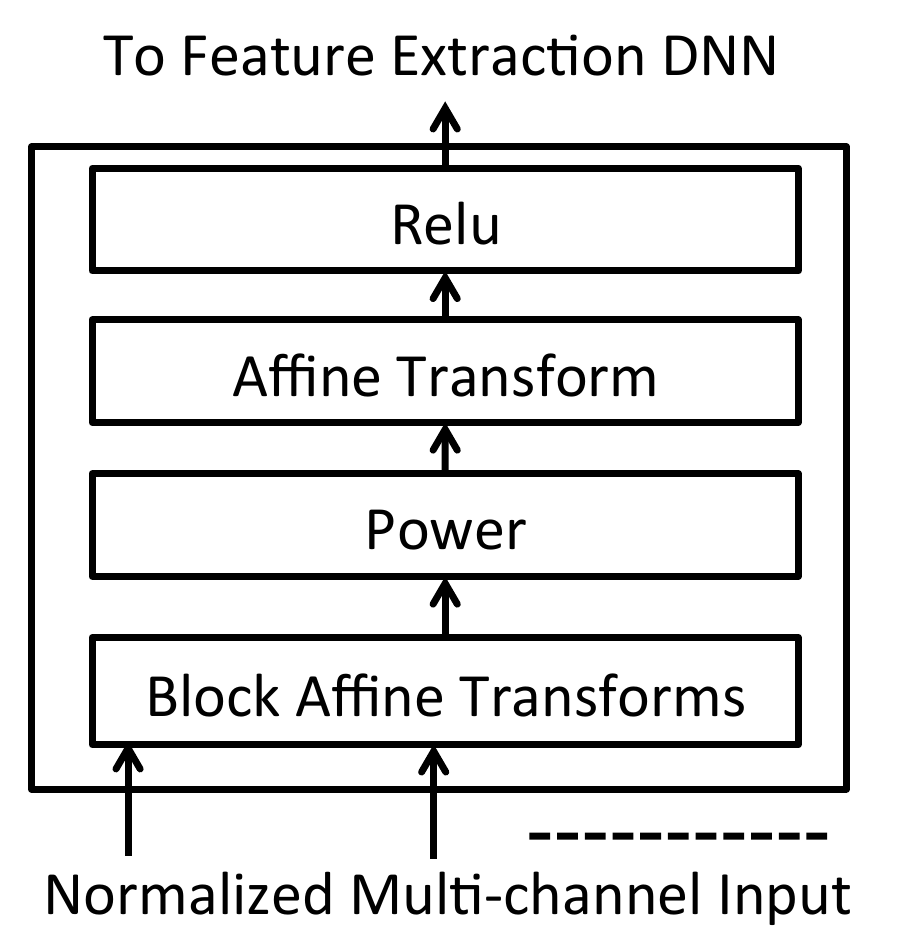}
\vspace{-0.75em}
\centering
\text{(c) Elastic SF }
\lb{fig:mcdnn3}
\end{minipage}
\caption{Our multi-channel (MC) networks}
\label{fig:all_mcdnn}
\end{minipage}
\begin{minipage}[c]{0.32\linewidth}
\includegraphics[width=\linewidth,left]{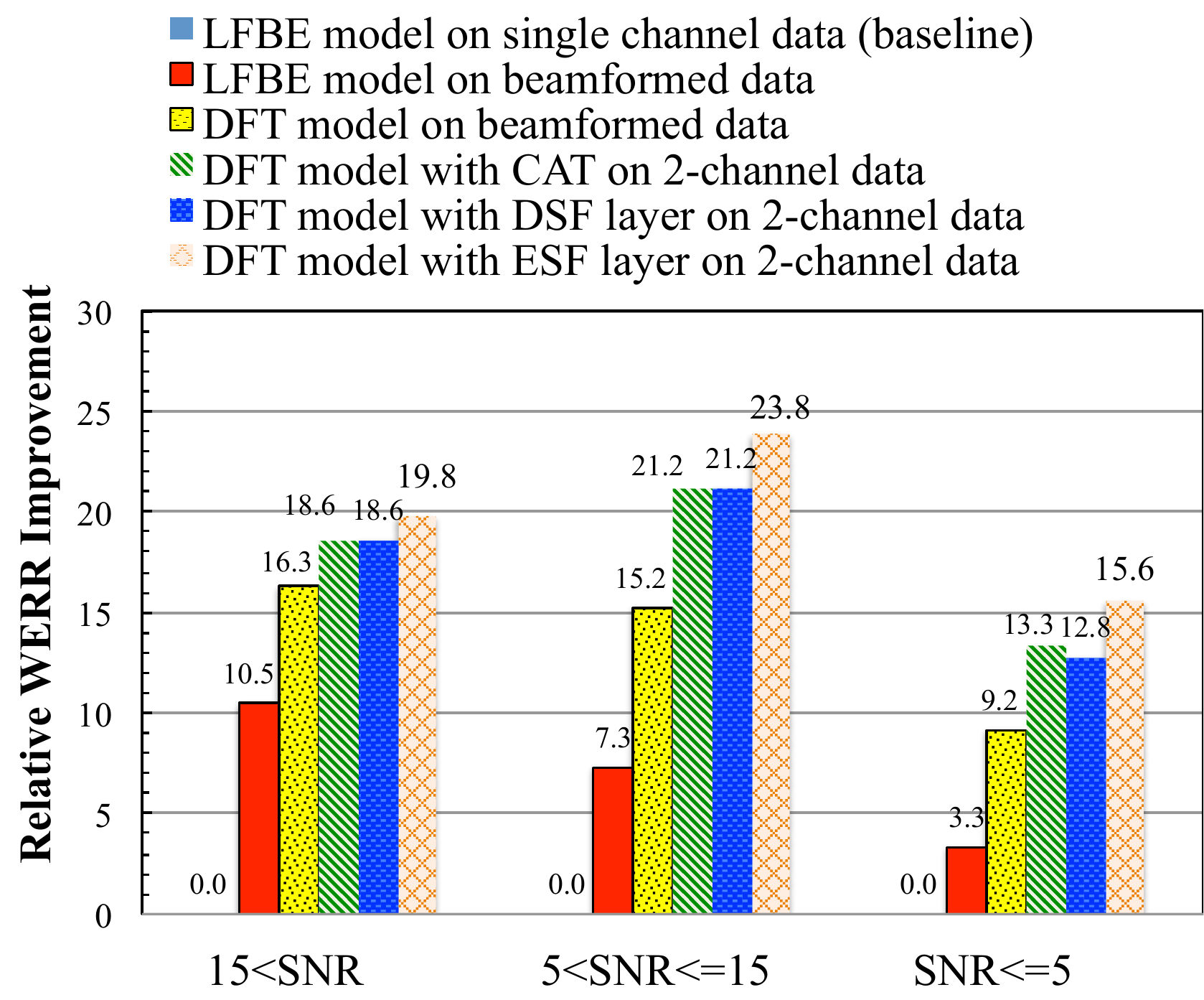}
\vspace{-2.2em}
\caption{Relative WERR of each method under different SNR conditions.}
\label{fig:WER_main}
\end{minipage}
\vspace{-0.5em}
\end{figure*}

In this work, we consider three types of MC network architectures as illustrated in figure~\ref{fig:all_mcdnn}. Figure~\ref{fig:all_mcdnn}~(a) depicts the simplest architecture considered in this work. In this structure, the concatenated multi-channel feature vector is transformed with a complex affine transform (CAT) followed by a complex square operation. This architecture is very similar to the complex linear projection model described in~\cite{LiSNCBMSSPCSWWV17} except for the bias vector. 

Figure~\ref{fig:all_mcdnn}~(b) shows another MC network architecture used in this work. The architecture is designed to model beamforming and beamformer selection. We initialize each row vector of the MC input layer with the SD beamformer's weights computed for different look directions. We then compute the pair-wise sum of squares of the output that corresponds to the power of the frequency component. The succeeding max-pooling operation is associated with beamformer selection based on the maximum power at each frequency bin. However, the deterministic nature of this output selection operation may result in an irrecoverable selection error. To alleviate this unrecoverable error, we allow the first spatial filtering layer to interact with different frequency components. In our preliminary experiment, providing this additional degree of freedom improved recognition accuracy. The output of the spatial filtering layer for each frequency $\omega_k$ can be obtained by taking the max value of the following sparse affine transform,
\vspace{-0.5em}
\begin{eqnarray*}
\text{pow} \left(
\begin{bmatrix}
         \bzero_{M(k-1)} \quad \bw^H _{\text{SD}} (\omega_k, \bp_1)  \quad \bzero_{M(K-k)} \\
                     \vdots \\
         \bzero_{M(k-1)} \quad \bw^H _{\text{SD}} (\omega_k, \bp_D) \quad \bzero_{M(K-k)} \\
\end{bmatrix}
\begin{bmatrix}
           \bX (\omega_1) \\
           \vdots \\
           \bX (\omega_K) \\
\end{bmatrix}
+ \bb
\right)
 \lb{eq:mcdnnb}
\vspace{-0.5em}
\end{eqnarray*}
where $ \bzero_L$ is $L$-dimension zero vector for initializing a non-target frequency weight to zero, $\bzero_0$ means null, $\bb$ is a bias vector and $\text{pow}()$ is the sum of squares of two adjacent values. As elucidated in section~\ref{sec:ex1}, initializing the first layer with beamformer's weight leads to a significant improvement in comparison to randomly initializing the weight matrix. 

Instead of deterministic selection of spatial layer's output, we consider another new network architecture that combines the weighted output power. Figure~\ref{fig:all_mcdnn}~(c) shows such an MC DNN architecture. This elastic MC DNN includes the block affine transforms initialized with SD beamformers' weights, signal power component, affine transform layer and ReLU. The output power of the spatial filtering layer is expressed with a block of frequency-independent affine transforms as
\vspace{-0.5em}
\begin{eqnarray*}
\begin{bmatrix}
           Y_1 (\omega_1) \\
           \vdots \\
           Y_D (\omega_1) \\
           \vdots \\
            Y_1 (\omega_K) \\
           \vdots \\
           Y_D (\omega_K) \\
\end{bmatrix}
= 
\text{pow} \left (
\begin{bmatrix}
         \bw^H _{\text{SD}} (\omega_1, \bp_1)  \bX (\omega_1) + \bb_{1\quad \quad \quad \quad} \\
         \vdots \\
         \bw^H _{\text{SD}} (\omega_1, \bp_D)  \bX (\omega_1) + \bb_{D \quad \quad \quad \quad} \\
         \vdots \\
         \bw^H _{\text{SD}} (\omega_K, \bp_1) \bX (\omega_K)  + \bb_{DK \quad \quad \quad} \\
         \vdots \\
         \bw^H _{\text{SD}} (\omega_K, \bp_D) \bX (\omega_K) + \bb_{D(K+1)} \\
\end{bmatrix}
\right ).
 \lb{eq:mcdnnc}
\vspace{-0.5em}
\end{eqnarray*}
In this elastic architecture, beamformer selection errors can be alleviated by combining the weighted output. Moreover, we can maintain the frequency independent processing constraint at the first layer, which leads to efficient optimization. These two points are the main differences between network architecture (b) and (c). 

In this paper, the MC network architectures of (a), (b) and (c) are referred as complex affine transform (CAT), deterministic spatial filtering (DSF) and elastic SF (ESF) network, respectively.
Notice that all the weights will be updated based on the cross entropy criterion. We expect that these MC networks will have the noise cancellation functionality by subtracting one spatial filtering output from another; this is learned from a large amount of data solely based on the ASR criterion instead of a hand-crafted adaptive manner. Similar to the permutation problem seen in the blind source separation, the MC network may permute different look directions for each frequency bin. However, the network should learn the appropriate weights for the senone classification task. 

\begin{figure*}[t]
\addtolength{\belowcaptionskip}{-3.0em}
\begin{minipage}[t]{0.32\linewidth}
\begin{center}
\centerline{\includegraphics[width=\linewidth,right]{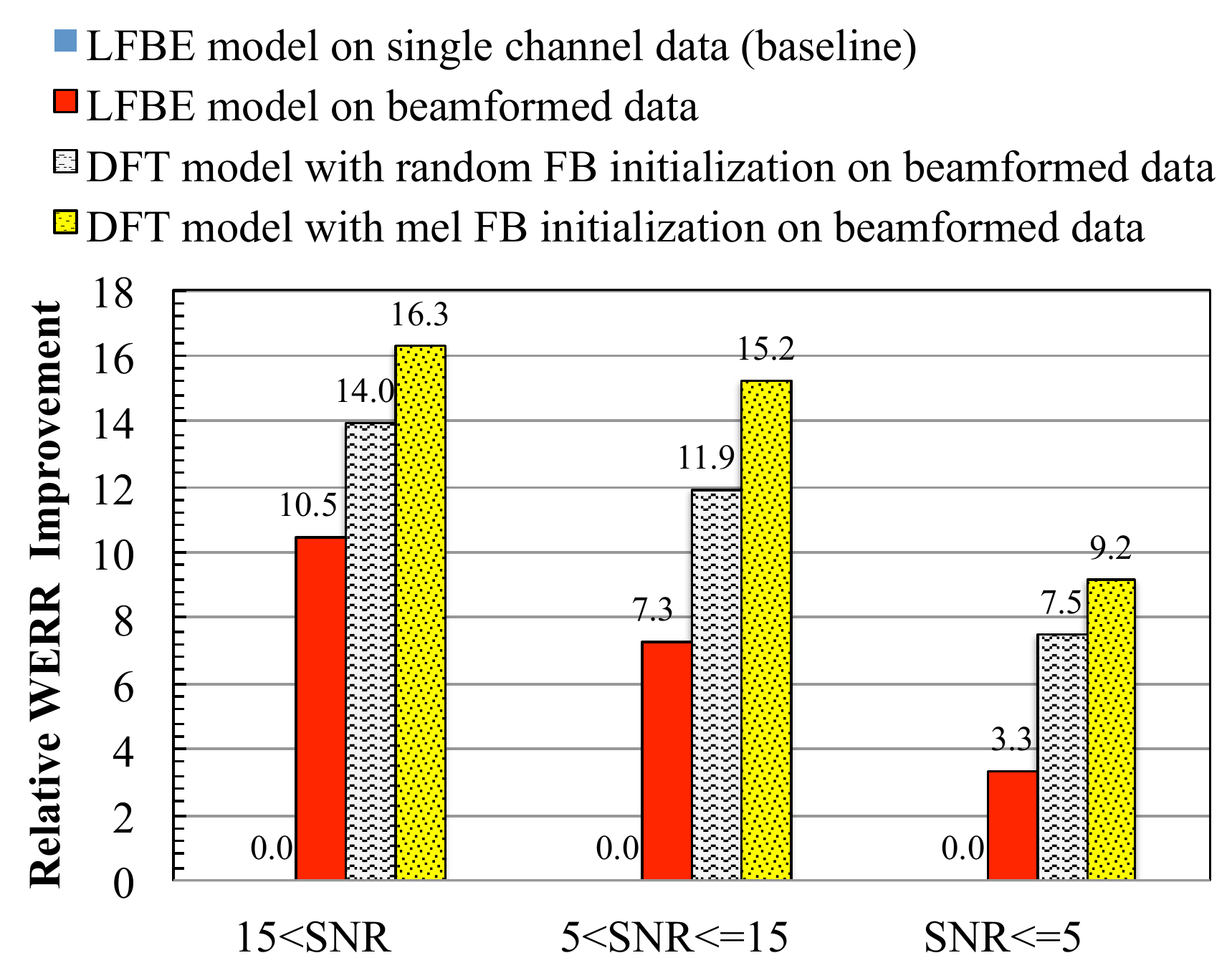}}
\vspace{-1em}
\caption{Comparison of feature front-ends}
\label{fig:WER_SDMs}
\end{center}
\end{minipage}
\begin{minipage}[t]{0.01\linewidth}
\includegraphics[width=\linewidth]{white.pdf}
\end{minipage}
\begin{minipage}[t]{0.32\linewidth}
\centerline{\includegraphics[width=\linewidth,left]{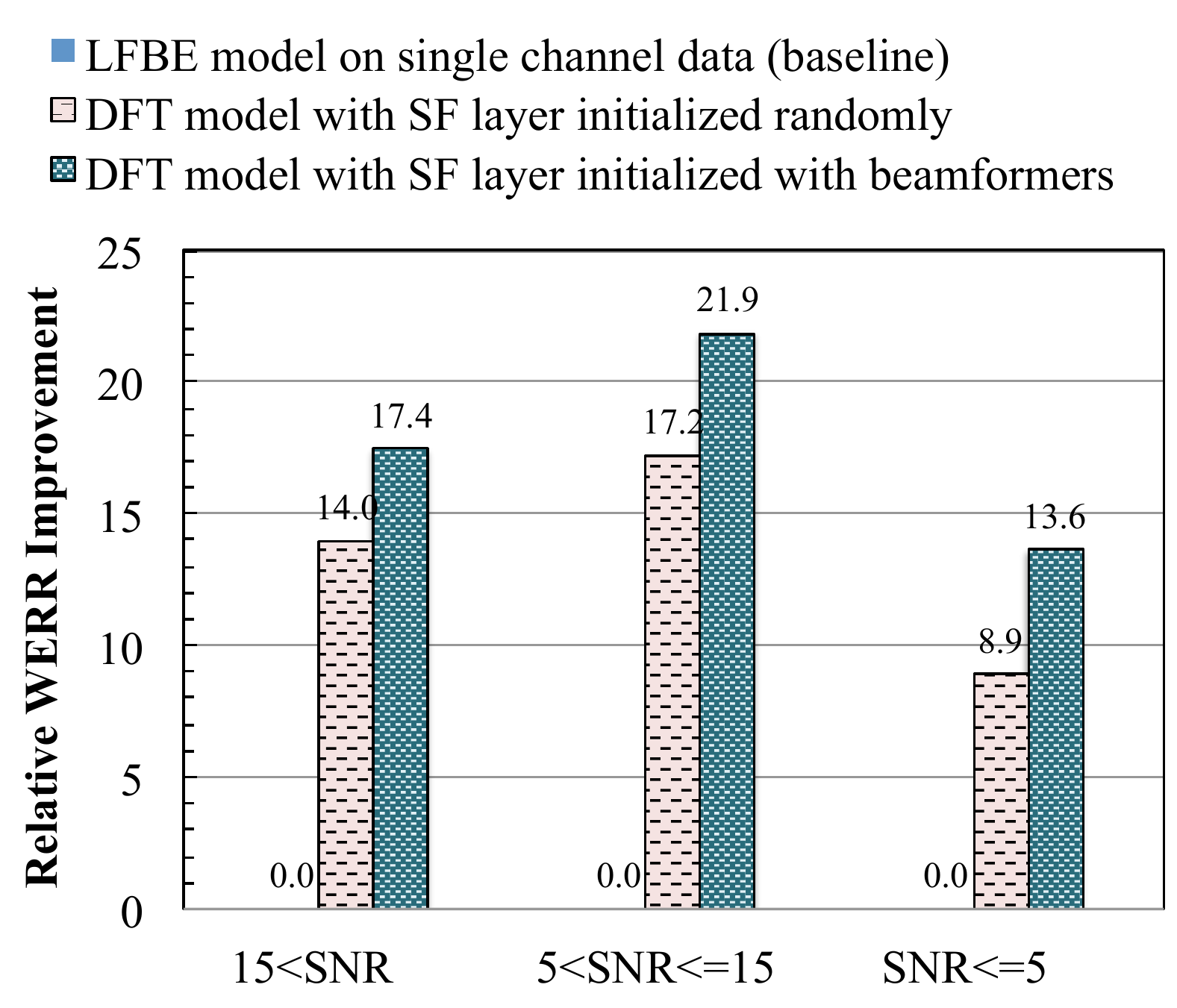}}
\vspace{-1em}
\caption{Effect of initialization of the SF layer.}
\label{fig:WER_BFinit}
\end{minipage}
\begin{minipage}[t]{0.01\linewidth}
\includegraphics[width=\linewidth]{white.pdf}
\end{minipage}
\begin{minipage}[t]{0.32\linewidth}
\centerline{\includegraphics[width=\linewidth, right]{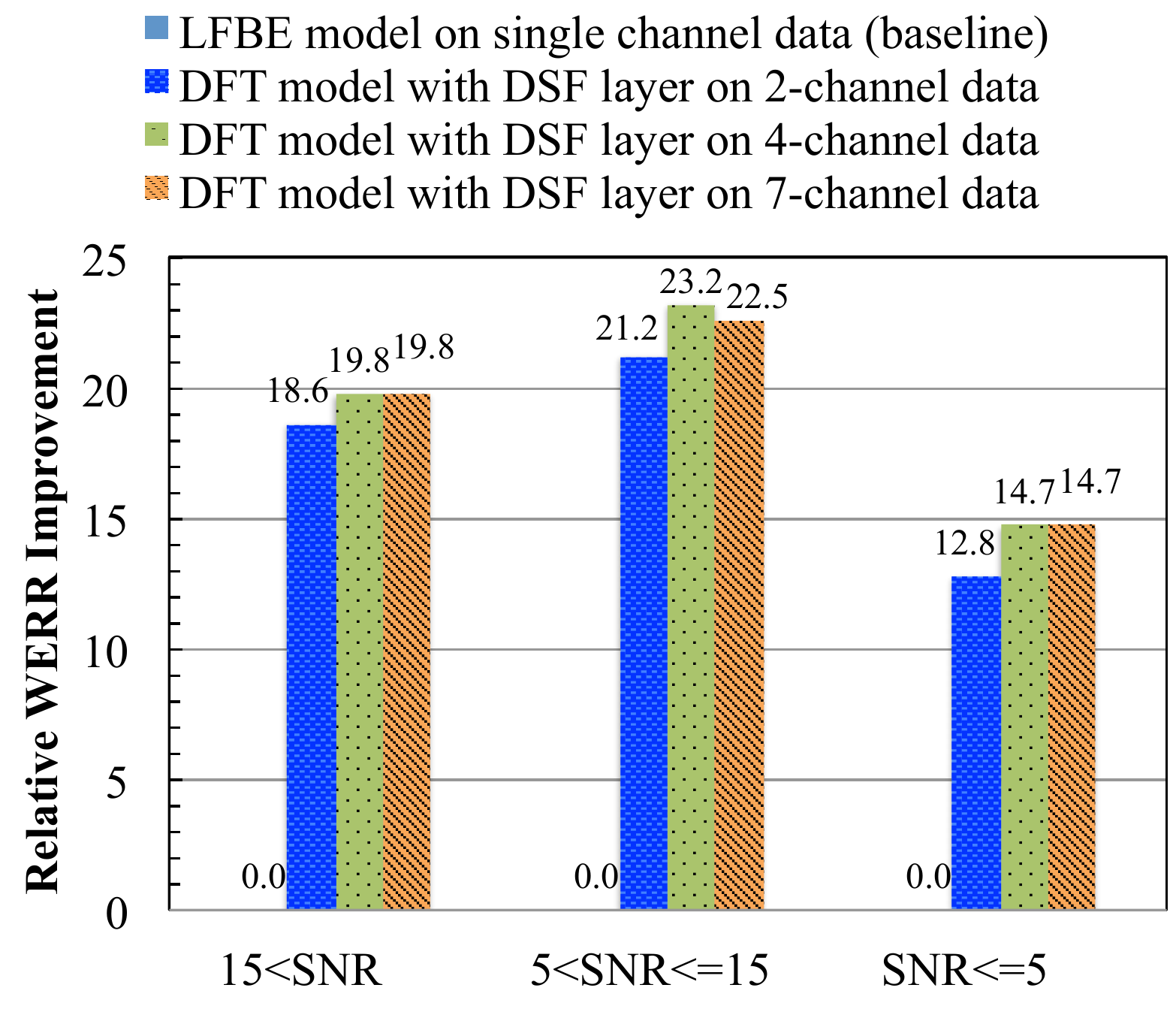}}
\vspace{-1em}
\caption{Relative WERR with respect to a number of microphones.}
\label{fig:WER_NoMics}
\end{minipage}
\vspace{-2em}
\end{figure*}
\vspace{-0.25em}
\section{ASR Experiment}
\vspace{-0.25em}
\label{sec:ex1}


In order to validate the efficacy of the MC acoustic modeling methods, we perform a series of ASR experiments using over 1100 hours of speech utterances from our in-house dataset. The training, test data amount to approximately 1,000 and 50 hours respectively. The device-directed speech data from several thousand anonymized users was captured using 7 microphone circular array devices placed in real acoustic environments. The interactions with the devices were completely unconstrained. Therefore, the users may move while speaking to the device. Speakers in the test set were excluded from the training set. 

As a baseline beamforming method, we use SD beamforming with diagonal loading adjusted based on~\cite{DocloM07}. The array geometry used here is an equi-spaced six-channel microphone circular array with a diameter of approximately 72 milli-meters and one microphone at the center. For beamforming, we used all the seven microphones. Multiple beamformers are built on the frequency domain toward different directions of interest and one with the maximum output energy is selected for the ASR input. It may be worth mentioning that conventional adaptive beamforming~\cite[S6,S7]{VanTrees2002} degraded recognition accuracy in our preliminary experiments due the difficulty of accurate voice activity detection and speaker localization on the real data. Thus, we omit results of adaptive beamforming in this work. For two channel experiments, we pick two microphones diagonally across the center of the circular array. Our experiments did not show sensitivity to microphone pair selection.  The baseline ASR system used a 64-dimensional LFBE feature with online causal mean subtraction (CMS)~\cite{King2017}. For our MC ASR system, we used 127-dimensional DFT coefficients removing the direct and Nyquist frequency components. The LFBE and FFT features were extracted every 10ms with a window size of 25ms and 12.5ms, respectively. Both features were normalized with the global mean and variances precomputed from the training data. The classification LSTM for the LFBE and FFT feature has the same architecture, 5 LSTM layers with 768 cells followed by the affine transform with 3101 outputs. 
All the networks were trained with the cross-entropy objective using a neural network toolkit~\cite{Strom15}. The Adam optimizer was used in all the experiments. For building the DFT model, we initialize the classification layers with the LFBE model. 


Results of all the experiments are shown as relative word error rate reduction (WERR) with respect to the performance of the baseline system. The baseline system is powerful enough to achieve a single digit number in a high SNR condition. The LFBE LSTM model for the baseline system was trained and evaluated on the single channel data captured with the center microphone of the circular array. The WERR results are further split by estimated signal-to-noise ratio (SNR) of the utterances. The SNR was estimated by aligning the utterances to the transcriptions with an ASR model and subsequently calculating the accumulated power of speech and noise frames over an entire utterance. 

Figure~\ref{fig:WER_main} shows the relative WERRs of the LFBE LSTM with the conventional 7-channel beamformer, the fully-trainable networks with the complex affine transform (CAT), deterministic spatial filtering (DSF) and elastic spatial filtering (ESF) layers. It is clear from figure~\ref{fig:WER_main} that SD beamforming with 7 microphones can provide better accuracy than the single channel system with the LFBE feature. It is also clear from figure~\ref{fig:WER_main} that the ASR accuracy can be further improved by jointly optimizing spatial filtering, feature extraction and classification LSTM layers. Moreover, figure~\ref{fig:WER_main} shows that the ESF network provides the best performance among three types of the MC networks optimized jointly. We consider this is because the ESF architecture can achieve better flexibility than the DSF network by linearly weighting the spatial filtering layer's output while imposing the frequency independent process on the input layer. These results also suggest that processing each frequency component independently at the first layer provides better recognition accuracy than combining them together.   

In addition to network architectures, we also explored the benefit of using a learnable feature extraction DNN. Figure~\ref{fig:WER_SDMs} compares the LFBE feature performance with the learnable LFBE network in the case that the whole networks are trained with the beamformed data. It is clear that the learnable feature extraction DNN alone can improve speech recognition performance. Initializing the filter-bank matrix with mel-filter bank coefficients leads to better accuracy than random initialization. Although we have not trained DNNs from numerous random initial points, our finding agrees with other literature~\cite{Bhargava,Tuske2016,Zeghidour2018}. 

Figure~\ref{fig:WER_BFinit} shows comparison between random and beamforming initialization. It is clear that initializing the first layer with beamformers' weights leads to better recognition performance. These results indicate that prior knowledge used in microphone array processing and speech recognition serves as good initialization. 

We also investigated ASR sensitivity as a function of a number of input channels. Figure~\ref{fig:WER_NoMics} shows the relative WERRs with respect to a number of microphones. We can see that there is a peak in ASR performance at four microphones, but the gain by using more than two microphones is small. This recognition performance saturation at three or four sensors is also observed in~\cite{Sainath2017,SwietojanskiSPL14,OchiaiICML17}.



\vspace{-0.25em}
\section{Conclusion}
\label{sec:conclusion}
\vspace{-0.25em}

We have proposed new MC acoustic modeling methods for DSR. The experiment results on the real far-field data have revealed that the fully learnable MC acoustic model with two-channel input can provide better recognition accuracy than the conventional ASR system, the LFBE model with 7 channel beamforming. It turned out that initializing the network parameters with beamformers' weights and filter bank coefficients led to better recognition accuracy. The result also suggests that it is important to have the structural prior at the first spatial filtering layer. Moreover, the experimental result on the beamformed data shows that the recognition accuracy can be further improved by updating the mel-filter bank parameter. We plan to scale up the training data size by combining multi-conditional training~\cite{raju2018data} and teacher-student semi-supervised learning method~\cite{Hari2019,Mosner2019}. 


\bibliographystyle{IEEEbib}
\bibliography{refs}

\begin{thebibliography}{10}

\bibitem{Omolog2001}
M.~Omologo, M.~Matassoni, and P.~Svaizer,
\newblock {\em Speech Recognition with Microphone Arrays}, pp. 331--353,
\newblock Springer Berlin Heidelberg, Berlin, Heidelberg, 2001.

\bibitem{Wolfel2009}
M.~W\"olfel and J.~W. McDonough,
\newblock {\em Distant Speech Recognition},
\newblock Wiley, London, 2009.

\bibitem{KumataniAYMRST12}
K.~Kumatani, T.~Arakawa, K.~Yamamoto, J.~W. McDonough, B.~Raj, R.~Singh, and
  I.~Tashev,
\newblock ``Microphone array processing for distant speech recognition: Towards
  real-world deployment,''
\newblock in {\em Proc. {APSIPA ASC}}, 2012.

\bibitem{KinoshitaDGHHKL16}
K.~Kinoshita, M.~Delcroix, S.~Gannot, E.~A.~P. Habets, R.~H\"ab{-}Umbach,
  W.~Kellermann, V.~Leutnant, R.~Maas, T.~Nakatani, B.~Raj, A.~Sehr, and
  T.~Yoshioka,
\newblock ``A summary of the {REVERB} challenge: state-of-the-art and remaining
  challenges in reverberant speech processing research,''
\newblock {\em {EURASIP} J. Adv. Sig. Proc.}, p.~7, 2016.

\bibitem{VirtanenBook2012}
T.~Virtanen, Rita Singh, and Bhiksha Raj,
\newblock {\em Techniques for Noise Robustness in Automatic Speech
  Recognition},
\newblock John Wiley \& {S}ons, West Sussex, UK, 2012.

\bibitem{Delcroix2014}
M.~Delcroix, T.~Yoshioka, A.~Ogawa, Y~Kubo, M~Fujimoto, N.~Ito, K.~Kinoshita,
  M.~Espi, T.~Nakatani, and Nakamura A.,
\newblock ``Linear prediction-based dereverberation with advanced speech
  enhancement and recognition technologies for the reverb challenge,''
\newblock in {\em Proc. of REVERB challenge workshop}, 2014.

\bibitem{Heymann2018}
J.~Heymann, M.~Bacchiani, and T.~Sainath,
\newblock ``Performance of mask based statistical beamforming in a smart home
  scenario,''
\newblock in {\em Proc. ICASSP}, 2018.

\bibitem{Fiscus2008}
J.~G. Fiscus and Jerome~Ajot ant J.~S.~Garofolo,
\newblock {\em The Rich Transcription 2007 Meeting Recognition Evaluation}, pp.
  373--389,
\newblock Springer Berlin Heidelberg, Berlin, Heidelberg, 2008.

\bibitem{Sullivan96}
T.~M. Sullivan,
\newblock {\em Multi-Microphone Correlation-Based Processing for Robust
  Automatic Speech Recognition},
\newblock Ph.D. thesis, Carnegie Mellon University, Pittsburgh, PA, 1996.

\bibitem{HimawanMS11}
I.~Himawan, I.~McCowan, and S.~Sridharan,
\newblock ``Clustered blind beamforming from ad-hoc microphone arrays,''
\newblock {\em {IEEE} Trans. Audio, Speech {\&} Language Processing}, vol. 19,
  no. 4, pp. 661--676, 2011.

\bibitem{McDonough2008}
J.~McDonough and M.~W\"olfel,
\newblock ``Distant speech recognition: Bridging the gaps,''
\newblock in {\em Proc. HSCMA}, 2008.

\bibitem{Seltzer2008}
M.~L Seltzer,
\newblock ``Bridging the gap: Towards a unified framework for hands-free speech
  recognition using microphone arrays,''
\newblock in {\em Proc. HSCMA}, 2008.

\bibitem{Seltzer2004}
M.~L. Seltzer, B.~Raj, and R.~M. Stern,
\newblock ``Likelihood-maximizing beamforming for robust hands-free speech
  recognition,''
\newblock {\em {IEEE} Transactions on Speech and Audio Processing}, vol. 12,
  no. 5, pp. 489--498, 2004.

\bibitem{Rauch2008}
B.~Rauch, K.~Kumatani, F.~Faubel, J.~W. McDonough, and D.~Klakow,
\newblock ``On hidden markov model maximum negentropy beamforming,''
\newblock in {\em Proc. IWAENC}, Sep. 2008.

\bibitem{Xiao16}
X.~Xiao, S.~Watanabe, H.~Erdogan, L.~Lu, J.~R. Hershey, M.~L. Seltzer, G.~Chen,
  Y.~Zhang, M.~I. Mandel, and D.~Yu,
\newblock ``Deep beamforming networks for multi-channel speech recognition,''
\newblock in {\em Proc. {ICASSP}}, 2016, pp. 5745--5749.

\bibitem{Sainath2017}
T.~N. Sainath, R.~J. Weiss, K.~W. Wilson, B.~Li, A.~Narayanan, E.~Variani,
  M.~Bacchiani, I.~Shafran, A.~Senior, K.~Chin, A.~Misra, and C.~Kim,
\newblock ``Multichannel signal processing with deep neural networks for
  automatic speech recognition,''
\newblock {\em IEEE Transactions on Speech and Language Processing}, 2017.

\bibitem{OchiaiICML17}
T.~Ochiai, S.~Watanabe, T.~Hori, and J.~R. Hershey,
\newblock ``Multichannel end-to-end speech recognition,''
\newblock in {\em Proc. {ICML}}, 2017.

\bibitem{SwietojanskiSPL14}
P.~Swietojanski, A.~Ghoshal, and S.~Renals,
\newblock ``Convolutional neural networks for distant speech recognition,''
\newblock {\em {IEEE} Signal Process. Lett.}, vol. 21, no. 9, pp. 1120--1124,
  2014.

\bibitem{Braun2018}
S.~Braun, D.~Neil, J.~Anumula, E.~Ceolini, and S.~Liu,
\newblock ``Multi-channel attention for end-to-end speech recognition,''
\newblock in {\em Proc. Interspeech}, 2018.

\bibitem{KimInterspeech16}
S.~Kim and I.~R. Lane,
\newblock ``Recurrent models for auditory attention in multi-microphone distant
  speech recognition,''
\newblock in {\em Proc. Interspeech 2016}, 2016, pp. 3838--3842.

\bibitem{FujimotInterspeech17}
M.~Fujimoto,
\newblock ``Factored deep convolutional neural networks for noise robust speech
  recognition,''
\newblock in {\em Proc. Interspeech}, 2017.

\bibitem{LiSNCBMSSPCSWWV17}
B.~Li et~al.,
\newblock ``Acoustic modeling for {G}oogle home,''
\newblock in {\em Proc. Interspeech}, 2017, pp. 399--403.

\bibitem{ErdoganInterspeech16}
H.~Erdogan, J.~R. Hershey, S.~Watanabe, M.~I. Mandel, and J.~Le Roux,
\newblock ``Improved {MVDR} beamforming using single-channel mask prediction
  networks,''
\newblock in {\em Proc. Interspeech}, 2016.

\bibitem{Higuchi2018}
T.~Higuchi, K.~Kinoshita, N.~Ito, S.~Karita, and T.~Nakatani,
\newblock ``Frame-by-frame closed-form update for mask-based adaptive {MVDR}
  beamforming,''
\newblock in {\em Proc. ICASSP}, 2018.

\bibitem{Haykin2001}
Simon~S. Haykin,
\newblock {\em Adaptive filter theory},
\newblock Prentice Hall, 2001.

\bibitem{DocloM07}
S.~Doclo and M.~Moonen,
\newblock ``Superdirective beamforming robust against microphone mismatch,''
\newblock {\em {IEEE} Trans. Audio, Speech {\&} Language Processing}, vol. 15,
  no. 2, pp. 617--631, 2007.

\bibitem{King2017}
B.~King, I.~Chen, Y.~Vaizman, Y.~Liu, R.~Maas, S.~Hari~Krishnan Parthasarathi,
  and B.~Hoffmeister,
\newblock ``Robust speech recognition via anchor word representations,''
\newblock in {\em Proc. Interspeech}, 2017.

\bibitem{Kumatani2017}
K.~Kumatani, S.~Panchapagesan, Minhua Wu, M.~Kim, N.~Str{\"{o}}m, G.~Tiwari,
  and A.~Mandal,
\newblock ``Direct modeling of raw audio with {DNN}s for wake word detection,''
\newblock in {\em Proc. {ASRU}}, 2017.

\bibitem{RichardSN13}
G.~Richard, S.~Sundaram, and S.~Narayanan,
\newblock ``An overview on perceptually motivated audio indexing and
  classification,''
\newblock {\em Proceedings of the {IEEE}}, vol. 101, no. 9, pp. 1939--1954,
  2013.

\bibitem{VanTrees2002}
H.~L. {Van Trees},
\newblock {\em Optimum Array Processing},
\newblock Wiley--Interscience, New York, 2002.

\bibitem{Strom15}
Nikko Str{\"{o}}m,
\newblock ``Scalable distributed {DNN} training using commodity {GPU} cloud
  computing,''
\newblock in {\em Proc. Interspeech}, 2015.

\bibitem{Bhargava}
M.~Bhargava and R.~Rose,
\newblock ``Architectures for deep neural network based acoustic models defined
  over windowed speech waveforms,''
\newblock in {\em Proc. Interspeech}, 2015.

\bibitem{Tuske2016}
Z.~T{\"{u}}ske, R.~Schl{\"{u}}ter, and H.~Ney,
\newblock ``Acoustic modeling of speech waveform based on multi-resolution,
  neural network signal processing,''
\newblock in {\em Proc. ICASSP}, 2016.

\bibitem{Zeghidour2018}
N.~Zeghidour, N.~Usunier, G.~Synnaeve, R.~Collobert, and E.~Dupoux,
\newblock ``End-to-end speech recognition from the raw waveform,''
\newblock in {\em Proc. Interspeech}, 2018.

\bibitem{raju2018data}
A.~Raju, S.~Panchapagesan, X.~Liu, A.~Mandal, and N.~Str{\"{o}}m,
\newblock ``Data augmentation for robust keyword spotting under playback
  interference.,''
\newblock {\em arXiv:1808.00563 e-prints}, Aug. 2018.

\bibitem{Hari2019}
S.~H.~K. Parthasarathi and N.~Str{\"{o}}m,
\newblock ``Lessons from building acoustic models from a million hours of
  speech,''
\newblock in {\em Proc. ICASSP}, 2019.

\bibitem{Mosner2019}
L.~Mo\v{s}ner, Minhua Wu, A.~Raju, S.~H.~K. Parthasarathi, K.~Kumatani,
  S.~Sundaram, R.~Maas, and B.~H{\"{o}}ffmeister,
\newblock ``Improving noise robustness of automatic speech recognition via
  parallel data and teacher-student learning,''
\newblock in {\em Proc. ICASSP}, 2019.

\end{thebibliography}

\end{document}